\documentclass{PoS}
\usepackage{capt-of}
\usepackage{epsfig}
\usepackage{amsmath}
\usepackage{mcite}

\newcommand{\be}{\begin{equation}}
\newcommand{\ee}{\end{equation}}
\newcommand{\bem}{\begin{minipage}}
\newcommand{\eem}{\end{minipage}}
\newcommand{\bet}{\begin{table}}
\newcommand{\eet}{\end{table}}

\title{Pion and $\rho$-meson form factors using four-point functions in N$_F$=2 QCD}

\ShortTitle{Pion and $\rho$-meson form factors using four-point functions in N$_F$=2 QCD}

\author{Constantia Alexandrou \\
        University of Cyprus, Department of Physics,
        P.O.Box 20537, 1678 Nicosia, Cyprus\\
        E-mail: \email{alexand@ucy.ac.cy}}

\author{\speaker{Giannis Koutsou}\\
        University of Cyprus, Department of Physics,
        P.O.Box 20537, 1678 Nicosia, Cyprus\\
        E-mail:\email{koutsou@ucy.ac.cy}}

\abstract{
Hadron wave functions and form factors can be extracted 
using four-point correlators. 
Stochastic techniques are used to estimate the all to all
propagators, which are required for the exact calculation 
of four-point functions. We apply 
the so called one-end trick to evaluate meson four-point functions.
We demonstrate the effectiveness of the technique in the case of the pion and
the $\rho$-meson where we extract  their charge distribution, 
as well as the form factors.
}

\FullConference{The XXV International Symposium on Lattice Field Theory\\
		 July 30-4 August 2007\\
		 Regensburg, Germany}

\begin{document}

\section{Introduction}
\vspace{-2pt}
The standard approach used in the evaluation of form factors 
in lattice QCD is to compute 
a three-point function. More detailed information on
hadron structure can be extracted from four-point correlators. 
 The quark distribution inside the hadron and hadron deformation are
just two such important aspects  that can be studied
using these correlators. 
The equal time density-density correlator provides 
a gauge invariant definition of the hadron 
"wave function" but  originally could only be evaluated approximately~\cite{PRD:66-094503}.
This is because four-point 
functions are harder to compute than two- and three-point functions,
 requiring the all to all propagator. 
The usual way to  estimate the all to all propagator is by 
employing stochastic techniques~\cite{PRD:58-034506}. 
In Ref~\cite{POS:2005-030,POS:2006-113} 
we used $Z(2)$ noise
combined with dilution to compute the all to all propagators
 and obtained results both for 
mesons and baryons~\cite{POS:2005-030,POS:2006-113}. 
In this work 
we generalize the so called one-end trick~\cite{PRD:73-074506},
originally devised as a method to 
calculate two-point functions,  to  evaluate four-point
 functions. We demonstrate that this
approach yields  more accurate results 
by evaluating the density-density correlator for the pion and $\rho$-meson 
and comparing the results to those
 obtained using standard stochastic 
techniques~\cite{POS:2005-030,POS:2006-113}. Furthermore,  
we extract the pion form factor  obtaining results
that have comparable errors as those obtained when 
one uses the one-end trick to compute the pion form factor using the 
three-point function~\cite{POS:2007-374}.
An advantage of using four-point functions is that we only need
one set of stochastic propagators to extract the form factor
for {\it any} momentum transfer unlike using three-point functions
where a new set is needed for every momentum.
We also show how to generalize our method  to other mesons and
give preliminary results on $G_1$, one of the three form factors 
of the $\rho-$meson. 

\vspace{-5pt}
\section{Four-point functions}
\vspace{-2pt}
Hadron four-point functions  are given by
\be
G^{j_\sigma}_h\left(\vec{x}_2,t_1,t_2\right)=\int d^3x_1d^3x\left<h(\vec{x},t)\right|j_\sigma^{q_f}(\vec{x}_2+\vec{x}_1,t_2) j_\sigma^{q_{f^\prime}}(\vec{x}_1,t_1)\left|h(\vec{x}_0,t_0)\right>
\label{Eq:Gen4pt}
\ee
where $j_\sigma^{q_f}$ is the normal ordered electromagnetic operator 
 $:\bar{q_f}\gamma_\sigma q_f:$ with $f$ being a flavor index, 
while $\left|h\right>$ denotes any hadronic state. 
The two integrations ensure zero momentum 
of the hadronic state; integrating over $\vec{x}_1$ sets the momentum 
of the source equal to that
of the sink and integrating over $\vec{x}$ sets both to zero. 
Thus to compute the four-point function
on the lattice, the all to all propagator from all sites $\vec{x}_1$ to 
 $\vec{x}$ is needed.

It is well known that an estimate for
the all to all propagator can be obtained using 
stochastic techniques \cite{PRD:58-034506}. In
brief, one inverts for a set of $N_r$ noise vectors obeying 
$\left<\xi^a_\mu(x)\xi^{b\dagger}_\nu(y)\right>_r=\delta(x-y)\delta_{a,b}\delta_{\mu,\nu}$ and $\left<\xi_\mu^a(x)\right>_r=0$
and estimates the all to all propagator by averaging the product of the 
solution vectors with the noise vectors over the stochastic ensemble.
Namely the quark propagator 
$G^{b,a}_{\nu,\mu}(x,y)\rightarrow \left<\phi^b_\nu(x)\xi^{a\dagger}_\mu(y)\right>_r$,
where $\xi$ is a noise vector and $\phi$ 
the solution vector.  
One, therefore, replaces every occurrence of $G$ with the product between 
$\xi$ and $\phi$ thereby  obtaining
the stochastic estimate for the four-point function. 
More details and results for hadron wave functions and the 
pion form factor obtained using this method can be found in 
Refs.~\cite{POS:2005-030,POS:2006-113},
where it is shown that with sufficient number of noise vectors 
and dilution one 
can obtain  a reasonable signal~\cite{CPC:172-145}.
We shall refer to this approach as Method I. 
Here we show how one can reduce stochastic noise by implementing 
the one-end trick~\cite{PRD:73-074506} for the computation of  meson
four-point functions. We shall  refer to this new
 approach as Method II.

\vspace{-5pt}
\section{Description of Method II}
The one-end trick was originally devised for the precise calculation 
of pion two-point functions. 
In its original form, one combines appropriately solution vectors 
so that an automatic 
summation over the source coordinate arises. 
Thus the number of stochastic inversions needed is 
reduced to a few  inversions,
 thereby suppressing stochastic noise. More explicitly, 
expanding the dot product between two solution 
vectors yields the pion two-point correlator summed over the source coordinate:
\be
\sum_{\vec{x}}\left<\phi^{\dagger a}_\mu(\vec{x},t)\phi^a_\mu(\vec{x},t)\right>_r = 
\sum_{\vec{x},\vec{y}_0,\vec{x}_0}\left<\left[G^{ab}_{\mu\nu}
(\vec{x},t;\vec{x}_0,t_0)\xi^b_\nu(\vec{x}_0,t_0)\right]^\dagger 
G^{ab^\prime}_{\mu\nu^\prime}(\vec{x},t;\vec{x}^{\prime}_0,t_0)\xi^{b^\prime}_{\nu^\prime}(\vec{y}_0,t_0)\right>_r \,,
\ee
where we assume that the noise vectors are localized on a certain time 
slice $t_0$. Taking the average of  
the noise vectors over the stochastic ensemble yields 
$\delta-$functions by definition. Thus we obtain
\be
\sum_{\vec{x},\vec{y}_0,\vec{x}_0}\left[G^{ab}_{\mu\nu}(\vec{x},t;\vec{x}_0,t_0)\right]^\dagger G^{ab^\prime}_{\mu\nu^\prime}(\vec{x},t;\vec{y}_0,t_0)\delta_{aa^\prime}\delta_{\nu\nu^\prime}\delta(\vec{x}_0-\vec{y}_0)
= \sum_{\vec{x},\vec{x}_0} Tr\left[\left|G(\vec{x},t;\vec{x}_0,t_0)\right|^2\right].
\label{Eq:PiOneEnd}
\ee
In the case of the pion Eq.~(\ref{Eq:PiOneEnd}) arises automatically
 since one combines the backward going propagator with
 the $\gamma_5$ pairs that appear 
in the pion interpolating operator. For a general interpolating field, 
say $\bar{q_i}\Gamma q_j$, 
where $\Gamma$ is any product of $\gamma$-matrices, 
the noise vectors must be diluted in spin i.e. 
$\xi^a_\mu(\vec{x}_0,t_0)_r=\xi^a(\vec{x}_0,t_0)\delta_{r\mu},\, r=1,...,4$. 
This imposes that
the number of noise vectors is in multiplets of four. 
In this case the solution vectors give 
$\phi^a_\mu(\vec{x},t;t_0)_r = \sum_{\vec{x}_0}G^{ab}_{\mu r}(\vec{x},t;\vec{x}_0,t_0)\xi^b(\vec{x}_0,t_0)$,
where the $t_0$ argument appearing in $\phi$ is to remind us
that the noise vector is localized on 
the time slice $t_0$. The combination given by
\be
\sum_{r,\vec{x}} \phi^a_\mu(\vec{x},t;t_0)_\nu(\Gamma\gamma_5)_{\nu r}\phi^{*a}_\kappa(\vec{x},t;t_0)_r(\gamma_5\bar{\Gamma})_{\kappa\mu}\,,
\ee
yields the two-point function of the meson summed over both sink and 
source coordinates. The downside
of this method is that, due to the automatic summation over both the source
and sink spatial coordinates, 
one cannot compute  two-point 
functions for arbitrary momenta using a given set of noise vectors. 
To utilize the one-end trick and extract the two-point correlator 
at finite momentum one must multiply 
the noise vectors with an appropriate phase prior to the inversion. 
Thus one needs $N_r$ 
inversions for every momentum vector.

The application of this method to four-point functions is appealing 
since by definition 
(see Eq.~(\ref{Eq:Gen4pt})) one is
interested in the case where the initial and final states are at rest. 
We show here how the
one-end trick can be implemented for the case of mesons. 
 The four-point function that we consider is
shown schematically in Fig.~\ref{Fig:GenT_sigma4pt}.  
At the propagator level we have 
\be
G^{j_\sigma}_\Gamma(\vec{x}_2;t_0,t_1,t_2,t) 
= \sum_{\vec{x}_1,\vec{x},\vec{x}_0} \left<\chi_\Gamma(\vec{x},t)\right|j^\sigma(\vec{x}_2+\vec{x}_1,t_2) j^\sigma(\vec{x}_1,t_1)\left|\chi_\Gamma(\vec{x}_0,t_, 0)\right>
\ee
$$
= \sum_{\vec{x}_1,\vec{x},\vec{x}_0} Tr\left[\gamma_5\gamma_\sigma G(\vec{x}_1,t_1;\vec{x}_0,t_0)      \bar{\Gamma}\gamma_5 G^\dagger(\vec{x}_2+\vec{x}_1,t_2;\vec{x}_0,t_0)  
                                             \gamma_5\gamma_\sigma G(\vec{x}_2+\vec{x}_1,t_2;\vec{x},t)     \Gamma\gamma_5 G^\dagger(\vec{x}_1,t_1;\vec{x},t) \right]
$$
where  $\bar{\Gamma}=\gamma_0\Gamma^\dagger\gamma_0$ and 
$\chi_\Gamma(x)$ is the interpolating field  of the meson that takes 
the general form $\bar{q}_i(x)\Gamma q_j(x)$ for $i\ne j$. Here we explicitly add a sum over the source coordinate $\vec{x}_0$ and we fix the \phantom{{\tiny blah~blah~blah~blah~blah~blah~blah~blah~blah~blah~blah~blah~blah}}

\hspace{-0.05\linewidth}\begin{minipage}{0.42\linewidth}
  \vspace{10pt}
  \scalebox{0.9}{
    \begin{tabular}{c|c|c|c}
      \hline
      \hline
      \multicolumn{4}{c}{$\beta$ = 5.6, $a^{-1}$ = 2.56(10) GeV}\\
      \hline
      \hline
      \# Confs. & $\kappa$ & am$_\pi$ & m$_\pi/$m$_\rho$\\
      \hline
      \multicolumn{4}{c}{$24^3\times40$ \cite{PRD:72-014503}}\\
      \hline
      185 & 0.1575 & 0.270(3) & 0.69\\
      150 & 0.1580 & 0.199(3) & 0.56\\
      \hline
      \multicolumn{4}{c}{$24^3\times32$ \cite{CPC:174-87}}\\
      \hline
      200 & 0.15825 & 0.150(3) & 0.45\\
      \hline
      \hline
    \end{tabular}
  }
  \captionof{table}{The simulation parameters used in our computations.}
  \label{Table:Confs}
  \vspace{10pt}
\end{minipage}
\hfill
\begin{minipage}{0.55\linewidth}
\vspace{-10pt}
time slice of the source, $t_0$, and the sink, $t$. The time slices, $t_1$ and $t_2$,
 where the currents are 
inserted, on the other hand, are free to 
take any value 
between $t_0$ and $t$. Thus one needs two sets of 
stochastic inversions, one set with the noise vectors localized 
on the time slice,  $t_0$ and one on the time slice, $t$.  One then  finds an appropriate combination of solution vectors 
such that the summation over source and sink coordinates
is carried out automatically. The combination: 
\end{minipage}

\be
\sum_{\vec{x}_1} Tr\left[\gamma_5\gamma_\sigma S(\bar{\Gamma};\vec{x}_1,t_1;\vec{x}_2+\vec{x}_1,t_2;t_0)\gamma_5\gamma_\sigma S(\Gamma;\vec{x}_2+\vec{x}_1,t_2;\vec{x}_1,t_1;t)\right]
\ee
\vspace{-5pt}
where $\,\,\,S^{ab}_{\mu\nu}(\Gamma;\vec{x}_2+\vec{x}_1,t_2;\vec{x}_1,t_1;t) = \sum_r \phi^a_\mu(\vec{x}_2+\vec{x}_1,t_2;t)_r(\Gamma \gamma_5)_{r\kappa}\phi^{*b}_\nu(\vec{x}_1,t_1;t)_\kappa\,\,\,$ achieves this.

\vspace{20pt}
\hspace{-20pt}\begin{minipage}{0.42\linewidth}
\hspace{5pt}\scalebox{0.45}{\input{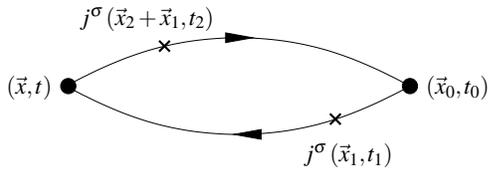}}
\center\captionof{figure}{The four-point function for mesons.}
  \label{Fig:GenT_sigma4pt}
\end{minipage}
\hfill
\begin{minipage}{0.55\linewidth}
  Throughout this work we use two degenerate flavors of dynamical  Wilson quarks
  \protect\cite{PRD:72-014503,CPC:174-87}. In all computations we
  employ Gaussian smearing combined with hypercubic (HYP) smearing~
  of the gauge
  links that  enter the Gaussian smearing function. The parameters
  of the Gaussian smearing are adjusted to  ensure
  minimal time evolution for filtering  the meson ground state. 
  The parameters 
  of our calculation are summarized in Table~\protect\ref{Table:Confs}. 
\end{minipage}

\vspace{-10pt}
\section{Meson wave functions}
The $\rho$-meson charge distribution is obtained using 
the equal time density-density correlator given by
\be
G^{j_0}_\rho\left(\vec{x}_1,t_1\right)=\int d^3x_2d^3x\left<\rho(\vec{x},t)\right|j_0^u(\vec{x}_2+\vec{x}_1,t_1) j_0^d(\vec{x}_2,t_1)\left|\rho(\vec{x}_0,t_0)\right> \,.
\ee
We test the new technique by comparing results for $G^{j_0}_\rho\left(\vec{x}_1,t_1\right)$
using Methods I and II.
In the large $t_1$ and $t-t_1$ limit when the $\rho$ state dominates,
 $G^{j_0}(\vec{x}_1, t_1) $, normalized over the spatial volume,
becomes time independent and it is denoted by $C_\rho(\vec{x}_1)$.
In the non - relativistic limit, this four-point function
reduces to the wave function squared. 
The ingredients needed in Method I are the point to all propagator from the
source and two all to all propagators
at time slices $t_1$ and $t$,  both of which  are kept fixed.
For this computation we use six sets of noise vectors 
diluted for each spin, color and even-odd sites i.e. we need 
$24 \times 6=144 $ inversions to obtain each stochastic propagator. 
This means   
a total of  $144\times 2+12 = 300$ inversions are required
for each gauge configuration \cite{POS:2006-113}. 
For Method II, on the other hand,
we used eight sets of spin diluted noise vectors at the source and sink thus 
a total of 64 inversions for each configuration. 
In Fig.~\ref{Fig:DDCompare} we show 
 a comparison between the results
obtained using Method I and II. 
What is plotted is the projection of the density-density~correlator,~$C_\rho(\vec{x}_1)$, along the spin axis taken to be the z-axis and perpendicular to it. The interpolating field used
for the $\rho$-meson is $\bar{u}\gamma_3 d$. 
As can be seen the statistical errors obtained when \phantom{blah blah blah blah blah blah blah blah blah blah blah blah}

\hspace{-20pt}\begin{minipage}{0.42\linewidth}
  \vspace{-70pt}

  \hspace{-10pt}\scalebox{0.95}{%
    \input{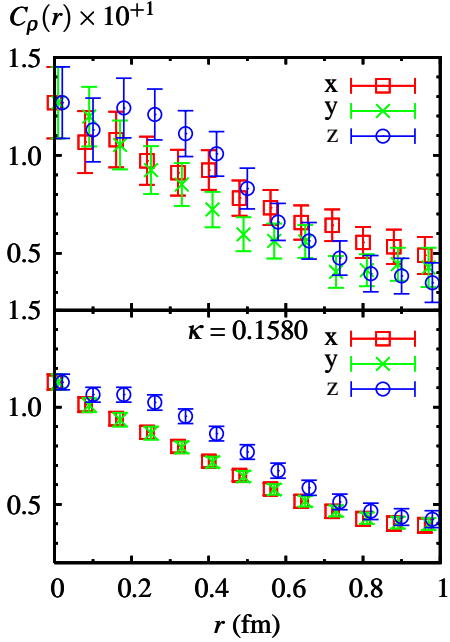}
  }
  \vspace{70pt}
\captionof{figure}{Projections of the $\rho$-meson density-density correlator
along the spin axis  and perpendicular to it.
Upper for Method I and lower for Method II at $\kappa=0.1580$.}
\label{Fig:DDCompare}
\end{minipage}
\hfill
\begin{minipage}{0.55\linewidth}
using Method II are almost four times smaller
despite the fact that we
use $144/32=4.5$  less number of noise vectors  to estimate the all to all
propagator. Therefore the 
improvement gained using the one-end trick is really significant,
reducing computational time by  two orders of magnitude.
The results obtained in Method II clearly reveal an
asymmetry in the charge distribution of the $\rho$-meson, which in Method I
was hard to see.

Having demonstrated the effectiveness  of Method II we use it, in what follows,
to study deformation in the $\rho$-meson as a function of the quark mass 
and to
extract the pion and $\rho$-meson form factors. 
In Fig.~\ref{Fig:ContAll} we show contour plots of the
density-density correlator of the $\rho$-meson, $C_\rho({\bf r})$,
 projected onto the $y$-$z$ plane.
As can be seen, for all three pion masses, we obtain an ellipse that
is elongated along the spin axis, showing
a clear deformation from
spherical symmetry. 
The corresponding contour plots for the pion show no deviation from
the circle as expected.
\end{minipage}


\begin{minipage}{0.9\linewidth}
\hspace{0\linewidth}
\scalebox{0.6}{
\input{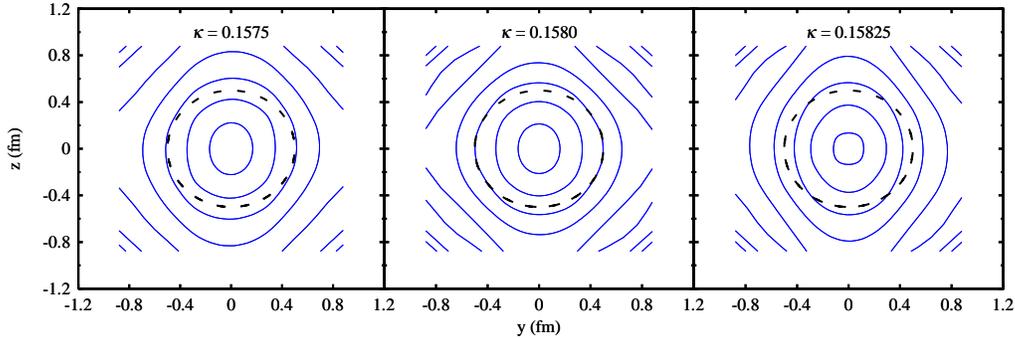}
}
\vspace{-7pt}
\captionof{figure}{
Contour plots of the charge distribution of the $\rho$-meson projected
onto the $y$-$z$ plane for all three $\kappa$ values studied. 
The dashed circles are to guide the eye.
}
\label{Fig:ContAll}
\end{minipage}


\section{Pion and $\rho$-meson form factors}
Form factors can be accurately extracted
using four-point functions by taking the Fourier transform 
of $G_h^{j_\sigma}(\vec{x},t_1,t_2)$ and allowing
large time separations between the
current insertions, $t_2-t_1$.
Therefore the extraction of form factors requires larger temporal extension
than the equal time density-density correlators. Methods to suppress
excited state contributions are therefore of crucial importance here.
Gaussian smearing combined with HYP smearing achieves  ground state dominance as
early as three time slices.
Taking the Fourier transform of the pion four-point correlator we obtain
$$
G_{\gamma_5}^{j_0}(\vec{p};t_1,t_2)\xrightarrow[t_1-t_0\gg1,\,t-t_2\gg1]{t_2-t_1\gg1}\left|\left<\chi_{\gamma_5}\right|\left.\pi(0)\right>\right|^2
\frac{\left|\left<\pi(0)\right|j_0\left|\pi(\vec{p})\right>\right|^2}{8m^2_\pi E(\vec{p})}e^{-E(\vec{p})(t_2-t_1)}e^{-m_\pi(t-(t_2-t_1)-t_0)}
$$
\be
= \left|\left<\left.\chi_{\gamma_5}\right|\pi(0)\right>\right|^2 \frac{\left|(E(\vec{p})+m_\pi)F_\pi(Q^2)\right|^2}{8m_\pi^2E(\vec{p})}e^{-E(\vec{p})(t_2-t_1)}e^{-m_\pi(t-(t_2-t_1)-t_0)} \quad,
\ee
where $F_\pi$ is the pion form factor and
$Q^2$ is the Euclidean momentum transfer squared.
The time dependencies and overlaps 
cancel by dividing with an appropriate combination of 
two-point functions:
\vspace{-5pt}
\be
R^{\,j^0}_{\gamma_5}(\vec{p};t_1,t_2) = \frac{\sqrt{4E(\vec{p})m_\pi}}{E(\vec{p})+m_\pi}\sqrt{\frac{G^{\, j^0}_{\gamma_5}(\vec{p};t_0,t_1,t_2,t) G_{\gamma_5}(\vec{p},t_1-t_0)}{G_{\gamma_5}(\vec{p},t_2-t_0) G_{\gamma_5}(\vec{0},t-(t_2-t_1)-t_0)}}
\ee
where $G_{\gamma_5}(\vec{p},t)$ is the pion two-point function at momentum $\vec{p}$. 
We search for a plateau of 
$R^{\,j^0}_{\gamma_5}(\vec{p};t_1,t_2)$ by varying the 
time difference $t_2-t_1$, as shown in Fig.~\ref{Fig:Plat}.
We perform the calculation for two source - sink separations, namely
$(t-t_0)/a=$14 and  16 to check 
that we have ground state dominance.
As can be seen, we obtain consistent plateau values
for both source - sink separations. 

\begin{minipage}{0.45\linewidth}
\hspace{-0.1\linewidth}
\scalebox{0.95}{\input{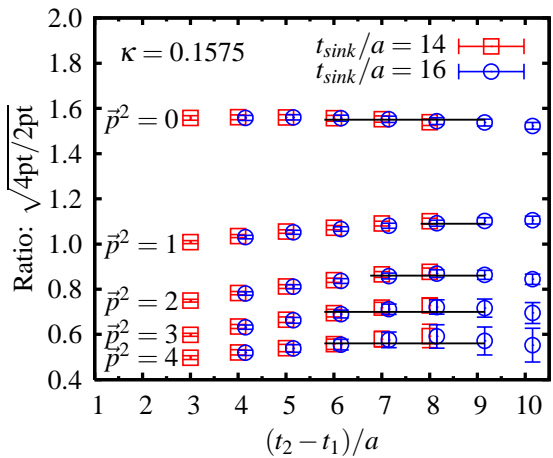}}
\vspace{-20pt}
\captionof{figure}{$R^{\,j^0}_{\gamma_5}(\vec{p};t_1,t_2)$ versus
$(t_2-t_1)/a$ for $\kappa=0.1575$. The 
range used for the fit is shown by the length of the lines.\newline 
\protect\phantom{blah blah blah blah blah }}

\label{Fig:Plat}
\end{minipage}
\hfill
\begin{minipage}{0.45\linewidth}
\hspace{-0.1\linewidth}
\scalebox{0.95}{\input{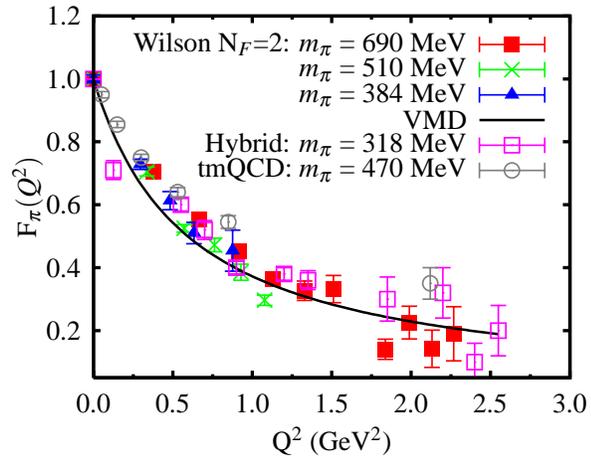}}
\vspace{-20pt}
\captionof{figure}{The pion form factor for three $\kappa$-values. We compare with results using the hybrid approach from \protect\cite{PRD:72-054506} and twisted mass results from \protect\cite{POS:2007-374}.}
\label{Fig:FFPi}
\end{minipage}
\vspace{10pt}


In Fig.~\ref{Fig:FFPi} we show the pion form factor for three $\kappa$-values 
compared with recent results obtained using three-point functions.
Results in the hybrid approach, that uses dynamical staggered sea
quarks and domain wall valence quarks, are obtained using sequential
inversions to compute the three-point function~\cite{PRD:72-054506}.
Results with dynamical twisted mass fermions, on the other hand,
use the one-end trick to compute the three-point function~\cite{POS:2007-374}. 
Our results compare very well to those obtained in the latter case, which
is closest to our approach. Assuming
vector meson dominance and taking  $m_\rho=0.77$~GeV 
we obtain the curve shown, for reference, in Fig.~\ref{Fig:FFPi}.

The $\rho$-meson has a Coulomb, $G_C$, a magnetic, $G_M$ and a quadrupole, 
$G_Q$ form factor. They can be parametrized in terms 
of $G_1$, $G_2$ and $G_3$ as
$$
G_Q=G_1 - G_2 + \left(1+\frac{Q^2}{4 M^2}\right)G_3,\qquad G_M=G_2,\qquad G_C=G_1 + \frac{2}{3}\frac{Q^2}{4 M^2}G_Q
$$
One can find combinations between initial and final $\rho$ polarizations 
and insertion directions ($\sigma$) that isolate individual form factors and for which decay to a pion
is forbidden.
Here we shall demonstrate the method
by showing preliminary results for $G_1$. 
As for the case of the pion form factor, $G_1$, can be extracted using only 
$j_0$:
\be
G^{\,j^0}_{\gamma_k}(\vec{p}_\perp;t_1,t_2) = \frac{\lambda^2(\vec{0})}{8M^2E(p_\perp)}\left(E(p_\perp)+M\right)^2G_1^2(Q^2)e^{-M(t-(t_2-t_1)-t_0)}e^{-E(p_\perp)(t_2-t_1)}
\ee

\hspace{-20pt}\begin{minipage}{0.485\linewidth}
\vspace{-10pt}
\hspace{-10pt}\scalebox{0.95}{\input{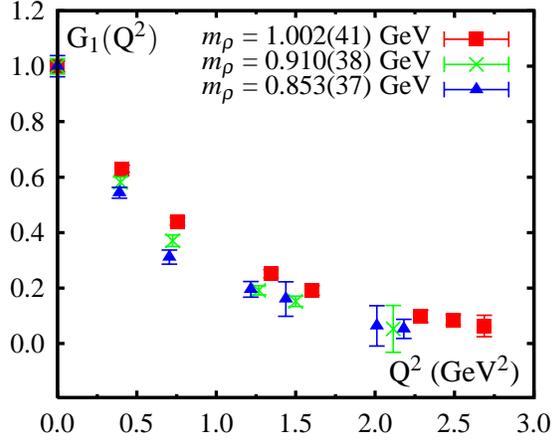}}
\vspace{-30pt}\captionof{figure}{$G_1$ as a function of the momentum transfer $Q^2$ for three $\kappa$ values.}
\label{Fig:RhoG1}
\end{minipage}
\hfill
\begin{minipage}{0.48\linewidth}
\vspace{-20pt}
where $\left<\Omega\right|\chi_{\gamma_k}\left|\rho(\vec{p},s)\right>=\lambda(\vec{p})\epsilon_k(\vec{p},s)$, $\sum_s\epsilon_k(\vec{p},s)\epsilon^*_{k^\prime}(\vec{p},s) = g_{kk^\prime} - \frac{ p_{k\phantom{^\prime}}  p_{k^\prime}}{M^2}$ 
and $\vec{p}_\perp$ is a momentum perpendicular to the $k$ direction.
As in the case of the pion form factor, we construct an appropriate
ratio and search
for a plateau in $t_2-t_1$. 
Results for $G_1$ are shown in Fig.~\ref{Fig:RhoG1}. 
They carry small statistical errors demonstrating the applicability of the method
for the extraction of the $\rho$-meson form factors. An analysis to extract all three form factors 
and subsequently derive physical quantities is in progress.
\end{minipage}

\section{Conclusions}
We have shown that the one-end trick can
be applied to evaluate accurately  four-point functions.
 Using this approach,  the density-density correlators are computed  
to sufficient accuracy  to show that the $\rho$-meson is deformed.
We also obtain accurate results for the pion form factor that 
compare favorably to the accuracy  obtained   
using  the one-end trick to compute the three-point function.
 The advantage of using four-point functions is
that  only one set of inversions is needed
for all momentum transfers, unlike in the case of three-point functions
where one needs new inversions for each value of the momentum transfer.
Preliminary results on the $\rho$-meson form factor, $G_1$, carry small statistical errors
demonstrating the applicability of the method also in the calculation of
the form factors of the $\rho$-meson.

\vspace{-4pt}
\bibliographystyle{JHEP_hack}
\bibliography{latt07.bib}
\end{document}